\documentclass[aps,prb,twocolumn,floatfix,superscriptaddress,%
amssymb,amsmath]{revtex4}

\usepackage{graphicx}

\begin{document}

\title{Selective optical manipulation of the spin state of a single magnetic impurity
in a semiconductor quantum dot}
\author{Alexander O. Govorov}
\affiliation{Department of Physics and Astronomy, Condensed Matter
and Surface Science Program, \\Ohio University, Athens, Ohio
45701-2979}
\author{Alexander V. Kalameitsev}
\affiliation{Institute of Semiconductor Physics, Novosibirsk,
630090 Russia}

\date{\today } 

\begin{abstract}

We describe the optical resonant manipulation of  a single
magnetic impurity in a self-assembled quantum dot. We show that
using the resonant pumping one can address and manipulate
selectively individual spin states of a magnetic impurity.
The mechanisms of resonant optical polarization of a single
impurity in a quantum dot involve anisotropic exchange
interactions and are different to those in diluted semiconductors.
A Mn impurity can act as qubit.  The limiting factors for the
qubit manipulation are the electron-hole exchange interaction and
finite temperature.

\end{abstract}

 \maketitle

The spins of electrons in semiconductors strongly couple with
electric and magnetic fields due to the spin-orbit and exchange
interactions. Spintronics and quantum computation utilize these
interactions to manipulate the electron spins
\cite{spinGeneral,QC}. One important class of spintronics
materials is diluted magnetic semiconductors \cite{Furdina} which
combine high-quality semiconductor structures with magnetic
properties of impurities. Since many semiconductors efficiently
emit and absorb light, the spin states of electrons and magnetic
impurities can be manipulated optically by using
circularly-polarized light pulses \cite{spin-optics1}. In diluted
magnetic semiconductors such as bulk crystals, quantum wells and
dots, photo-generated excitons interact with a large collection of
spins of Mn impurities and therefore a large number of degrees of
freedom becomes involved
\cite{spin-optics2,Yakovlev,MnBulkOptics,SpinPoralBulk,Kulakovskii-PRB,magnetic
dots,Cincinnati}. In these systems, it is challenging to address
individual spins of Mn atoms. At the same time, the quantum
computational schemes are based on qubits, pairs of
well-controlled quantum states. These elementary blocks, qubits,
should be made interacting or decoupled on demand. In a diluted
magnetic semiconductor, even a single Mn atom has 6 spin states
($I_{Mn}=5/3$).  Therefore, 15 different pairs of states (qubits)
can be defined for a single Mn impurity. Here we study a system
which allows us to manipulate optically a single Mn spin. This
system is composed of a quantum dot (QD) and a single Mn impurity.
Note that the optical properties of a QD with a single Mn impurity
were recently discussed in refs.~\cite{MnspinQD,Govorov-PRB204}.

This letter describes a single Mn impurity embedded into a
self-assembled QD. Importantly, such a system permits efficient
selective optical control and manipulation of individual spin
states and defining a single qubit for the Mn impurity. This
ability comes from the exciton spectrum of a QD with a Mn atom. An
exciton in a QD has a well-defined discrete spectrum and,
simultaneously, strongly interacts with the Mn spin via the
exchange interaction. Since the exciton and Mn spin functions
become strongly mixed, the resonant optical excitation strongly
affect the spin state of Mn impurity. In particularly, we show
that one can {\it write} spin states of Mn atom.
Since spin relaxation of paramagnetic ions in the absence of
carriers (i.e. after the exciton recombination) is an extremely
slow process ($\sim 10~\mu s$), a single Mn spin is a very
promising candidate for spintronics applications. The mechanisms
of Mn-spin polarization in a QD are qualitatively different to
those in bulk materials because of the discrete character of
quantum states. In bulk, the photo-generated spin-polarized
electrons transfer their spin to the Mn atoms or induce an
effective magnetic field which polarizes the impurity system
\cite{SpinPoralBulk}. In a QD, the spin orientation of Mn atom
comes from the three-body interactions involving an electron,
hole, and Mn spin. The ability to manipulate a pair of chosen
states (qubit) comes from the resonant excitation of a certain
spin state of the exciton-Mn system. In addition, we describe the
specific optical signatures of a Mn atom embedded into a QD. In
contrast to the undoped self-assembled quantum dots, the optical
emission of a laterally-asymmetric quantum dot becomes {\it
circularly polarized} due to the exciton-Mn interaction.

We now consider a model of disk-shaped self-assembled QD taking
into account only the heavy holes (HH) states in the valence band.
The QD potential strongly confines the electron and HH envelope
wave functions, $\phi_e({\bf r}_e)$ and $\phi_h({\bf r}_h)$ and
the exchange interactions in the exciton determine the spin state
of exciton.  According to the conventional model, the Mn-hole and
Mn-electron exchange interactions are proportional to $\delta({\bf
r}_{e(h)}-{\bf R}_{Mn})$, where ${\bf R}_{Mn}$ is the Mn position
and ${\bf r}_{e(h)}$ is the electron (hole) coordinate. Then, the
spin Hamiltonian takes the form

\begin{eqnarray}
\hat{H}_{spin}=\hat{H}^{exc}_{Mn-hole}+ \hat{H}^{exc}_{Mn-e}+
\hat{H}^{exc}_{e-hole},  \label{HamiltExciton}
\end{eqnarray}
which includes three types of exchange interaction. The
anisotropic exchange interaction between the Mn spin and HH is
$\hat{H}^{exc}_{Mn-hole}=\frac{\beta}{3}| \phi_h({\bf
R}_{Mn})|^2\hat{j}_{h,z}\hat{I}_{Mn,z}=A_h\hat{j}_{h,z}\hat{I}_{Mn,z}$;
the Mn-electron interaction is isotropic,
$\hat{H}^{exc}_{Mn-e}=\alpha|\phi_e({\bf R}_{Mn})|^2\hat{{\bf
s}}\hat{{\bf I}}_{Mn}=A_e\hat{{\bf s}}\hat{{\bf I}}_{Mn}$. Here,
$\hat{{\bf I}}_{Mn}$ and ${\bf \hat{s}}$ are the Mn and electron
spins, respectively ($s=1/2$, $I_{Mn}=5/2$); $\hat{{\bf j}}$ is
the HH momentum ($j_h=3/2$ and $j_{h,z}=\pm3/2$). The anisotropic
e-HH interaction is given by the operator \cite{bookIv}:

\begin{eqnarray}
\hat{H}^{exc}_{e-hole}=
\sum_{i=x,y,z}{a_i\hat{j}_i\hat{s}_i+b_{i}\hat{j}^3_i\hat{s}_i},
\label{ehExc}
\end{eqnarray}
where $a_i$ and $b_i$ are constants. The eigenstates of the
Hamiltonian (\ref{HamiltExciton}) are linear combinations of 24
functions, $|I_{Mn,z},j_{h,z},s_{e,z}>$, where $I_{Mn,z}$,
$j_{h,z}$, and $s_{z,e}$ are the z-components of the corresponding
momenta.

\begin{figure}[tbp]
\includegraphics*[width=0.5\linewidth, angle=90]{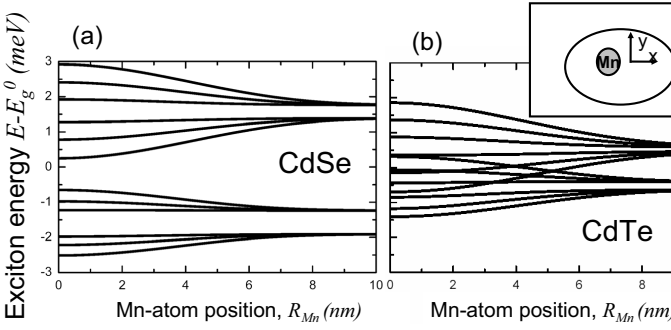}
\caption{Exciton energy spectrum as a function of the
$x$-coordinate of the Mn-atom; $R_{Mn,y}=0$ and $R_{Mn,z}=L_z/2$.
The energy $E_g^0$ is the band gap of a QD. (a) corresponds to
$b_x=-0.7$, $b_y=-0.2$, $b_z=-0.4$, and $a_z=-1.2~meV$; (b)
corresponds to $-0.23,\ -0.066,\ -0.4$, and $-0.13~ meV$. Insert:
geometry of the system.} \label{f1}
\end{figure}

The strength of the Mn-hole and Mn-electron exchange interaction
depends on the position of the Mn impurity with respect to the QD
(fig.~\ref{f1}, insert) and is given by the coefficients $A_h$ and
$A_e$. For the parameters of exchange interaction, we choose
$\alpha N_0=0.29~eV$ and $\beta N_0=-1.4~eV$, typical numbers for
$II-VI$ materials \cite{Kulakovskii-PRB,MnspinQD}; here $N_0$ is
the number of cations for unit volume. In fig.~\ref{f1}, we show
the calculated spectrum of exciton as a function of the Mn
position for two set of parameters $a_i$ and $b_i$. The first set
(fig.~\ref{f1}a) relates to the case of CdSe QD
\cite{FineStructure} and the second (fig.~\ref{f1}b) corresponds
to the weaker e-hole exchange interaction and to a CdTe QD. For
the QD wave function, we use a convenient approximation:
$\phi_{e(h)}({\bf r})=B_{e(h)}sin(\frac{\pi z}{L_z})
e^{-x^2/l^2_{x,e(h)}-y^2/l^2_{y,e(h)}}$ \cite{AxelAPL,PRB}. The
sizes of QD are $l_{x,e}=l_{x,h}=6~nm$, $l_{y,e}=l_{y,h}=4~nm$,
and $L_z=2~nm$. In fig.~\ref{f1}, we observe that, in the limit
$R_{Mn}\rightarrow\infty$, a QD has a spectrum determined by the
laterally-anisotropic e-hole exchange interaction; the
photoluminescence (PL) spectrum in this case is linearly polarized
along the X and Y directions \cite{FineStructure}. For the case
$R_{Mn}\sim l_{dot}$, the energy structure strongly changes due to
the Mn-induced exchange interaction that has cylindrical, $D_{2d}$
point-group symmetry. This cylindrical symmetry results in the
nonzero spin polarization of the wave functions.

\begin{figure}[tbp]
\includegraphics*[width=0.8\linewidth, angle=90]{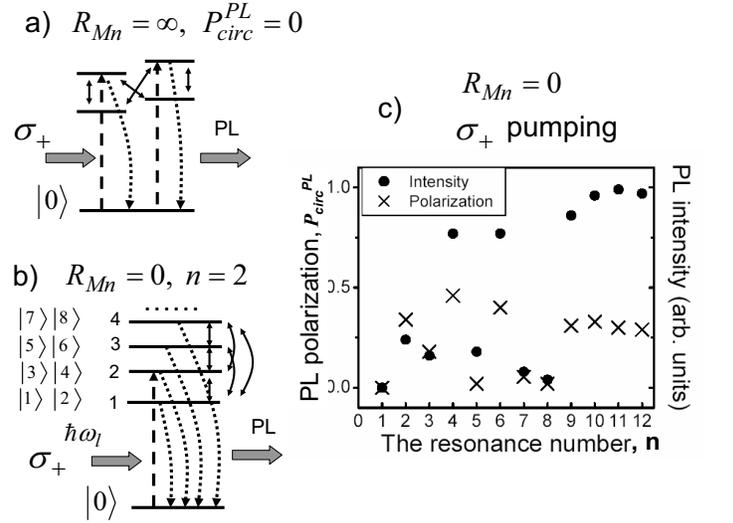}
\caption{(a) Exciton spectrum of a QD without a Mn impurity.
Arrows show the pumping, emission, and relaxation processes. (b)
The pumping and relaxation processes in the presence of a Mn atom;
the second level is optically excited, $n=2$ and
$\hbar\omega_l=E_3=E_4$. (c) Calculated degree of circular
polarization and PL intensity for $\sigma_+$ pumping in a QD with
$R_{Mn}=0$.} \label{fig2}
\end{figure}

The excitonic wave functions $|\gamma>$ of energies $E_\gamma$
($\gamma=1,2,...24$) are doubly degenerate (fig.~\ref{f1}) and
their energy spectrum consists of 12 energy levels, $n=1,2,...12$
(fig.~\ref{fig2}b). Wave functions in a pair of degenerate states
can be written using the two non-crossing subspaces. For example,
the states $|1> (|2>)$ are composed of wave functions with
$I_{Mn,z}+j_{h,z}+s_{e,z}=2m+1/2$
($I_{Mn,z}+j_{h,z}+s_{e,z}=2m+1+1/2$), where $m$ is integer.

{\it Master equation.} The pumping and PL processes are described
by the master equation,

\begin{eqnarray}
\frac{\partial \hat{\rho}}{\partial
t}=\frac{i}{\hbar}[\hat{\rho},\hat{H}_0+\hat{V}_{opt,+}(t)]+L(\hat{\rho})
\label{DM},
\end{eqnarray}
where $\hat{\rho}$ is the density matrix,
$\hat{V}_{opt,+}(t)=W_0(\hat{p}_+e^{i\omega_l t}
+\hat{p}_-e^{-i\omega_l t})$ is the interaction with classical
circularly-polarized light, $\hat{p}_\pm=\hat{p}_x\pm i\hat{p}_y$,
and $\omega_l$ is the laser frequency. $L(\hat{\rho})$ is the
relaxation operator within the Markovian approximation:
\begin{eqnarray}
[L(\hat{\rho})]_{\gamma,\gamma'}=-\frac{\Gamma_\gamma+\Gamma_{\gamma'}}{2}\hat{\rho}_{\gamma,\gamma'}
\ (\gamma \neq \gamma') \nonumber
\end{eqnarray}
\begin{eqnarray}
[L(\hat{\rho})]_{\gamma,\gamma}=-\Gamma_\gamma\hat{\rho}_{\gamma,\gamma}+
\sum_{\gamma''}\hat{\rho}_{\gamma'',\gamma''}\Gamma_{\gamma''\rightarrow\gamma},
\label{MApprox}
\end{eqnarray}
where $\Gamma_\gamma=\Gamma^{intra}_\gamma+\Gamma^{rad}_\gamma$,
$\Gamma^{intra}_\gamma$ is the rate if intra-band relaxation of an
exciton $\gamma$ (this relaxation involves both spin and energy),
$\Gamma^{rad}_\gamma$ is the inter-band, radiative rate. The
latter can be written as $\Gamma^{rad}_\gamma=B_\gamma\Gamma_0$,
where $\Gamma_0=1/\tau_0^{rad}$ is the rate of radiative
relaxation for the bright heavy-hole exciton in a QD without a Mn
impurity; the coefficient $B_\gamma$ depends on the spin
configuration of the exciton.

The intra-band relaxation rates
$\Gamma_{\gamma\rightarrow\gamma'}$ between different momentum and
spin states of an exciton depend on a particular system and come
from the spin-orbit, electron-phonon, and strong exchange
interactions in the conduction and valence bands. In undoped II-VI
QDs, the spin relaxation time of an exciton is typically longer
that the radiative time. For example, it was found in
ref.~\cite{MicherPRL} that the spin relaxation time is longer that
$0.5~ns$. Ref.~\cite{hole} reports the hole relaxation time is of
order of $10~ps$. For our calculations, we take
$\Gamma_1=1/5~ns=0.2~ns^{-1}$. For the radiative life-time, we
take a typical value $\tau^{rad}_0=0.5~ns$.  At finite temperature
$T$, the rates  $\Gamma_{\gamma\rightarrow\gamma'}$ depend on the
energy separation $E_{\gamma'\gamma}=E_{\gamma'}-E_{\gamma}$. Here
we will use a simplified model,
$\Gamma_{\gamma\rightarrow\gamma'}=\Gamma_1$ if
$E_{\gamma'\gamma}<0$ and
$\Gamma_{\gamma\rightarrow\gamma'}=\Gamma_1e^{-E_{\gamma'\gamma}/k_BT}$
if $E_{\gamma'\gamma}>0$. This model describes also the
thermally-activated transitions. The long spin relaxation time of
paramagnetic ions comes from the spin-lattice interaction
($\tau_{spin-lattice}\sim1-100~ms$) \cite{Mn relaxation}. For the
Mn relaxation rate in the absence of an exciton, we choose
$\Gamma_{Mn}=1/10~ms$.

If the light intensity in the pulse is relatively low, the
equations (\ref{MApprox}) are reduced to a system of rate
equations for the diagonal components
$\rho_{\alpha}=\rho_{\alpha,\alpha}$, where $\alpha$ can be an a
exciton-Mn state $|\gamma>$ or a state without exciton,
$|I_{Mn,z}>$. In the following, we will be solving numerically a
system of rate equations for the two cases. As for the equilibrium
density matrix, $\rho^0_{I_{Mn,z},I_{Mn,z}}=1/6$ and otherwise
zero.

{\it Circularly-polarized optical emission}. The emission
intensity of a photo-generated exciton $\gamma$ is proportional
to: $I_{\pm,\gamma}=(2\pi/\hbar)\sum_{I_{Mn,z}}|<I_{Mn,z}|{\hat
V}_{\pm}^{PL}|\gamma>|^2$, where $|\gamma>$ is the initial exciton
state, $|I_{Mn,z}>$ are the final states of the system, and ${\hat
V}_{\pm}^{PL}=V_0(\hat{p}_x\mp i\hat{p}_x)$ are the operators for
the photon emission in the direction $+z$. Then, the degree of
circular polarization of a given exciton state $\gamma$ can be
calculated as
$P_{circ,\gamma}=(I_{+,\gamma}-I_{-,\gamma})/(I_{+,\gamma}+I_{-,\gamma})$.
In the limit $R_{Mn}\rightarrow\infty$ ($A_{e(h)}\rightarrow0$),
the exciton wave functions are determined by the anisotropic
e-hole exchange interaction, the emission spectrum is linearly
polarized \cite{FineStructure}, and $P_{circ,\gamma}=0$. For small
$R_{Mn}$, the calculated degrees $P_{circ,\gamma}$ for excitons in
a CdTe QD are close to $\pm1$ because $A_{h}>a_i,b_i$. Therefore,
the Mn-related symmetric exchange interaction for small $R_{Mn}$
determines the optical response of excitons.

Since individual excitons have a nonzero degree of circular
polarization, PL can also be circularly polarized under the
circularly-polarized pumping. We now consider the resonant
$\sigma_+$-optical pumping of CdTe QD into a given pair of states
with the energy $E_\gamma$ ($\hbar\omega_l=E_{\gamma})$
(fig.~\ref{fig2}b). Also, we assume a short laser pulse with a
duration $\Delta t\ll\tau_{rad}$.
The integrated PL intensity of a QD is given by the diagonal
components of the density matrix: $I_{PL,\pm}=\int dt(\sum_\gamma
\rho_{\gamma,\gamma}\Gamma^{PL,\pm}_\gamma)$, where
$\Gamma^{PL,\pm}_\gamma$ are the probabilities to emit $\pm$
photons in the vertical direction.

In the limit $R_{Mn}\rightarrow\infty$, the upper bright states
can be pumped optically (fig.~\ref{fig2}a). Since the individual
excitons are linearly polarized, the degree of circular
polarization
[$P_{circ}^{PL}(\hbar\omega_l=E_{\gamma})=(I_{PL,+}-I_{PL,-})/(I_{PL,+}+I_{PL,-})$]
is zero.

In the case of $R_{Mn}<\infty$, the time-integrated PL intensity
becomes nonzero. In fig.~\ref{fig2}, we show the time-integrated
PL intensity $I_{PL}=I_{PL,+}+I_{PL,-}$ and $P_{circ}^{PL}$ for
all resonant excitations $n=1,2,...12$. The resonance number $n$
in fig.~\ref{fig2}c is the number of the exciton energy level
counting as shown in fig.~\ref{fig2}b. For example, the resonance
$n=1$ relates to the states $\gamma=1,2$, etc. Overall, the
circular polarization of PL under circular pumping is well
expressed. The circular polarized PL signal was recently recorded
in semi-magnetic CdTe QDs \cite{Cincinnati}. This observation is
in contrast to the previous studies of undoped II-VI and II-V QDs
\cite{FineStructure}. Our theory suggests a qualitative
explanation for this observation.


\begin{figure}[tbp]
\includegraphics*[width=0.8\linewidth, angle=90]{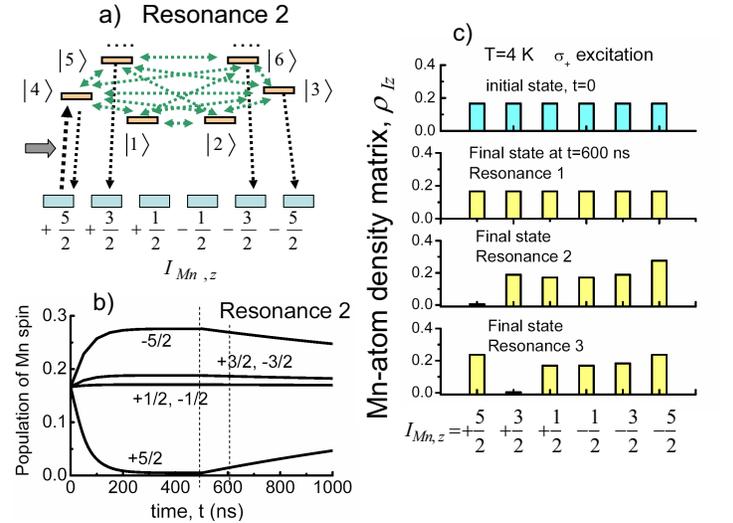}
\caption{(a) Schematics of pumping, relaxation, and PL processes
in a QD with Mn atom; the second level is resonantly excited
$\hbar\omega_l=E_4=E_3$; here we show only 6 lowest states. (b)
Calculated population of Mn atom states as a function of time for
the pulse $\Delta t=0.5~\mu s$. Dashed lines indicate the end of
pulse and the measurement time. (c) Population of Mn spin states
at $t=0.6~\mu s$ for different resonant pumping. The initial spin
state of Mn is randomized. } \label{fig3}
\end{figure}

\begin{figure}[tbp]
\includegraphics*[width=0.8\linewidth, angle=90]{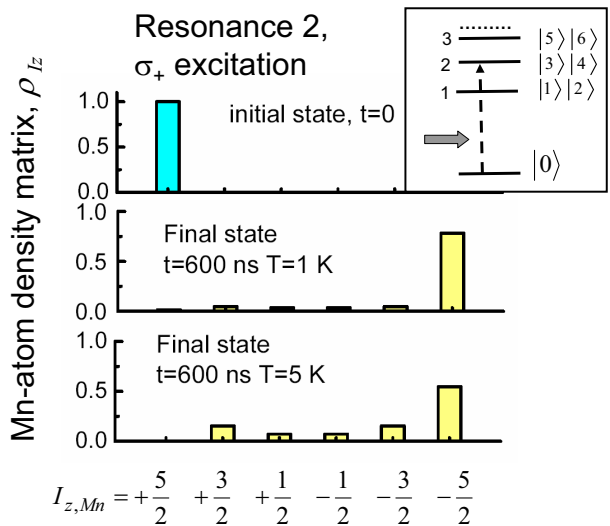}
\caption{Population of Mn spin states at $t=0.6~\mu s$ for the
resonant pumping into the second level (insert) at different
temperatures. The initial spin state of Mn is $+5/2$.}
\label{fig4}
\end{figure}

{\it Optical writing and manipulation of the Mn spin}. We now
calculate the time evolution of Mn spin in the presence of and
after a long $\sigma_+$ pulse of small intensity ($\Delta
t=0.5~\mu s$, $P_{cv}^2W_0^2/\hbar^2\Gamma_{rad}^2=0.02$). In the
initial state at $t=0$, the Mn atom is in thermal equilibrium,
$\rho_{I_{Mn,z}}=1/6$. The evolution of Mn spin strongly depends
on how the QD is pumped. As an example, consider the resonance
$n=2$ ($\hbar\omega_l=E_3=E_4$) and pumping into the exciton
states $|3>$ and $|4>$ (fig.~\ref{fig3}a,b). The states $|3>$ and
$|4>$ optically couple mostly with the Mn states
$I_{Mn,z}=\pm5/2$. Therefore, the $\sigma_+$ pulse affects mostly
the Mn spin state $I_{Mn,z}=5/2$. We see that, with increasing
time, the probability $\rho_{I_{+5/2}}$ decreases, while
$\rho_{I_{-5/2}}$ increases. In this process, the $\sigma_+$ pulse
excites the state $|4>$, then the exciton $|4>$ makes transition
to the state $|3>$, and finally the exciton $|3>$ recombines,
contributing to the Mn state $I_{Mn,z}=-5/2$. Fig.~\ref{fig3}
shows the Mn state at the "measurement" time $t=0.6~\mu s$. In
this way, the Mn spin population becomes non-equilibrium. The
exciton ground states $|1>$ and $|2>$ are mostly dark and do not
play an important role; they trap and release excitons to the
upper energy levels at final $T$.

The optical probing ({\it reading}) of the state of Mn atom in a
QD can be made with the absorption or PLE single-dot
spectroscopies which are presently available \cite{KhaledAPL}. The
absorption spectrum of a QD is very sensitive to the Mn spin state
\cite{MnspinQD,Govorov-PRB204}. For the $\sigma_\pm$ photons with
energy $E_3=E_4$ ($n=2$), the light absorption intensity is mostly
proportional to $\rho_{I_{\pm5/2}}$ and thus the change in
$\rho_{I_{\pm5/2}}$ can be recorded.

The non-equilibrium spin distribution $\rho_{I_{Mn}}$ strongly
depends on the resonance pumping. The resonant pumping into the
ground states $|1>$ and $|2>$ is not efficient since these states
are mostly dark (fig.~\ref{fig3}c). The resonant pumping $n=3$
(fig.~\ref{fig3}c) results in a decrease of $\rho_{I_{Mn,+3/2}}$
and an increase of $\rho_{I_{Mn,-3/2}}$. Although, the states
$\pm5/2$ become also involved. By using proper resonances, most of
the spin Mn states can be addressed in this way. The resonant
pumping of a given exciton state can be combined with {\it
Hanbury-Brown/Twiss (HBT) setup} \cite{HBT}. Consider weak
non-resonant illumination of the QD and assume that the QD emits a
$\sigma_+$ photon with the energy $E_4$ at $t=0$. It means that
the QD state after the emission process is the Mn state
$I_{Mn,z}=+5/2$. This state can then be used as an initial state
in our scheme with the polarized laser pulse started at $t=0$. In
the presence of the laser pulse, the Mn state $I_{Mn,z}=+5/2$
turns into the state $I_{Mn,z}=-5/2$ (fig.~\ref{fig4}). At the
measurement time $t=0.6~\mu s$, the Mn spin is mostly in the state
$I_{Mn,z}=-5/2$. The temperature effect leads to the mixing with
other exciton states and more Mn spin states become involved
(fig.~\ref{fig4}). For the qubit operation in this particular QD,
it is logical to choose the Mn spin states $\pm5/2$ and the
resonance $E_4$ ($n=2$) since, for this case, the mixing with the
other Mn states is minimum. The preparation of qubit in the state
$I_{Mn,z}=+5/2$ can be done by detecting a $\sigma_+$ photon with
the energy $E_4$. Then, the rotation of the qubit
($+5/2\rightarrow -5/2$) can be realized with a polarized laser
pulse. The limiting factors are finite temperature and anisotropic
electron-hole exchange interaction which create mixing between the
qubit and the other states of the Mn spin. The next logical step
is to involve two Mn atoms and consider a two-qubit regime.

To conclude, we have studied the optical properties of the spin
state of a single magnetic impurity embedded into a semiconductor
QD. It is shown that the Mn impurity can be optically manipulated
and the impurity spin can act as a qubit.

The author would like to thank Pierre Petroff for motivating
discussions. This work was supported by Ohio University and AvH
Foundation.

\end{document}